\documentclass[11pt]{article}
\usepackage{amsmath,amsthm,latexsym,amssymb,amsfonts}

\newtheorem{lemm}{Lemma}[section]
\newtheorem{prop}[lemm]{Proposition}

\newtheorem{defi}[lemm]{Definition}
\newtheorem{assu}[lemm]{Assumption}

\newcommand{\scripta}{\mathfrak{A}} 
\newcommand{\hilb}{\mathcal{H}}
\newcommand{\Hilb}{\mathfrak{h}}
\newcommand{\one}{\text{\bf 1}}
\newcommand{\E}{\mathcal{E}}
\newcommand{\abar}{\overline{\mathcal{A}}}
\newcommand{\isom}{\cong}
  
\DeclareMathOperator{\porder}{\mathcal{P}}
\DeclareMathOperator{\im}{Im}
\DeclareMathOperator{\re}{Re} 
\DeclareMathOperator{\Span}{span}
\DeclareMathOperator{\cyl}{Cyl}
\DeclareMathOperator{\dom}{dom}  
\DeclareMathOperator{\sutwo}{SU(2)}
\newcommand{\rep}[1]{\pi\left(#1\right)}
\newcommand{\comm}[2]{\left[#1\,,\,#2\right]}
 
\newcommand{\scpr}[2]{\left\langle#1\,,\, #2 \right\rangle} 

\newcommand{\norm}[1]{\left\lVert #1 \right\rVert}     
\newcommand{\betr}[1]{\left\lvert #1 \right\rvert}

\begin{document}
\title{Some Results Concerning the Representation Theory of the Algebra
  Underlying Loop Quantum Gravity}
\author{
Hanno Sahlmann\thanks{sahlmann@apctp.org. This work was done while 
at the MPI f\"ur Gravitationsphysik, Albert-Einstein-Institut, Potsdam, Germany}\\
{\small Asia Pacific Center for Theoretical Physics, Pohang (Korea)}}

\date{{\small PACS No. 04.60.Pp, 11.10.Cd, Preprint AEI-2002-056}}

\maketitle

\begin{abstract}
Important characteristics of the loop approach to quantum gravity 
are a specific choice of the algebra $\scripta$ of (kinematical) 
observables and 
of a representation of $\scripta$ on a measure space over the space
of generalized connections. This representation is singled out by its 
elegance and diffeomorphism covariance. 

Recently, in the context of the quest for \textit{semiclassical
states}, states of the theory in which the quantum gravitational
field is close to some classical geometry, it was realized that it
might also be worthwhile to study different representations of the
algebra $\scripta$.   

The content of the present work is the observation that under some
mild assumptions, the mathematical structure of representations of 
$\scripta$ can be analyzed rather effortlessly, to a certain extent: 
Each representation can be labeled by sets of functions and measures 
on the space of (generalized) 
connections that fulfill certain conditions.  
\end{abstract}
\section{Introduction}
Loop quantum gravity (LQG for short) is a promising approach to the
problem of finding a quantum theory of gravity, and has led to many
interesting insights (for reviews see \cite{Rovelli:1997yv,Thiemann:2007zz}). It is based on the formulation of gravity as a
constrained canonical system in terms of the Ashtekar variables
\cite{Ashtekar:1986yd}, a canonical pair of an SU(2)-connection (in its
real formulation) and a triad field.

One of the interesting features of LQG (and perhaps one reason for its 
success) is its specific choice of basic variables: The configuration 
variables are holonomies along curves in the spacial slices of the
spacetime, the basic momentum variables are integrals of a triad field 
over surfaces in the spacial slices of the spacetime. 
This is in contrast to ordinary quantum field theories, where both the 
configuration and the momentum observables are three dimensional
integrals of the basic field and its conjugate momentum. 
The choice of basic variables in LQG is, however, well motivated since 
in contrast to other possibilities, these variables can be
defined without recurse to a fixed classical background geometry, and
it furthermore leads to well defined operators for interesting
geometric quantities such as area and volume. 

A quantum theory for this type of basic variables was first given by Rovelli
and Smolin in \cite{Rovelli:1989za}. Since then, much work has gone
into extracting the essence of this quantization and putting it onto 
firm mathematical ground. Key ideas in this context were the use of 
$C^*$-algebraic methods \cite{Ashtekar:1991kc} and projective limit techniques
\cite{Ashtekar:1994mh,Ashtekar:1994wa} resulting in what is now called the 
\textit{connection representation}. This representation is based on a
Hilbert space which is an $L^2$-space over the space of connections
with respect to a certain measure, the Ashtekar-Lewandowski measure. 
The holonomies act as multiplication operators and the integrated 
triad fields as certain vector fields. 
Due to its diffeomorphism invariance and mathematical elegance, this
representation is considered the fundamental representation of LQG.  

That it might nevertheless be interesting to also consider 
representations other than the AL-representation was realized when
attempts were made to construct states for LQG in which the
quantum gravitational field behaves almost classical. 
The first proposal in this direction was contained in 
\cite{Arnsdorf:1999wn}. There, the goal was to find states for LQG
that have semiclassical properties for spacetimes with non-compact
spacial slices.   
Representations that are inequivalent to the AL-representation also 
seem to arise if one implements the ideas \cite{Bombelli:2000ua} about
the use of statistical geometry for the construction of semiclassical
states. 
Finally, in a series of works \cite{Varadarajan:2001nm,Varadarajan:1999it,Ashtekar:2001xp}, 
measures on the space of
generalized connections were constructed that derive from the Gaussian
measure of ordinary (background dependent) free quantum field theory. 

The representation theory for the holonomy algebra is well understood
and many representations inequivalent to the AL-representation have 
been considered in the literature. 
Less attention has been paid to the question of what happens when one
also takes the integrated triads into consideration. 
The main observation of the present work is that due to the 
structure of its commutation relations, representations of the
combined algebra of holonomies and integrated triads can, without
effort, be analyzed to a certain extent: 
Each representation can be labeled by sets of functions and measures 
on the space of (generalized) connections that fulfill certain
conditions. 

Since this article was first published as a preprint, numerous further works on the representation theory, and on representations of the algebra of LQG have been published \cite{Okolow:2003pk,Sahlmann:2003in,Varadarajan:2007dk,Dziendzikowski:2009rv} culminating in a uniqueness theorem for diffeomorphism invariant cyclic representations \cite{Lewandowski:2005jk}. Many of these works are more or less directly based on the analysis contained in the present one. 

Finally,  a cautionary remark is in order. The considerations of the
present work are mostly of mathematical nature. Truly interesting,  
albeit difficult, tasks would be to state physically 
motivated criteria for singling out interesting classes of
representations, actually constructing such representations, and
understanding the physical content of representations constructed by 
mathematical considerations. Little of this will be addressed in this
work. However, we hope that it can be used as a starting point 
when approaching those questions motivated by physics. 

To finish this introduction, we should mention that the occurrence of
inequivalent representations of the observable algebra is well known
from quantum field theory and quantum statistical mechanics
\cite{Haag:1992hx}: In that context, it was realized
that the choice of representation for the observable algebra contains 
important physical information: Roughly speaking, whereas the
algebraic structure of the theory encodes the physical
system one is considering, the chosen representation carries the
global information about the physical state the system is in. It
might for example decide whether the system is in a ground or in a
thermal state or whether the state carries a global charge.  
Since the change of the global properties of a state of the system 
is not always physically realizable (it might necessitate an infinite
amount of energy or the creation of charges) the emergence of inequivalent
representations is quite natural. 
The consideration of representations different from the
AL-representation in the quest for the semiclassical
regime of LQG fits quite nicely into this general picture.   
\section{LQG briefing}
It is sometimes useful to quantize
a given classical system in two steps. The first consists in
associating to each member of a chosen set of classical observables,
an operator in some (abstract) $*$-algebra $\scripta$, such that
\begin{itemize}
\item The Poisson structure of the classical observables is mirrored
  as closely as possible by the commutators within the algebra
  ("Poisson brackets go to commutators").
\item Complex conjugate on classical observables are mapped to
  conjugates under the $*$-operation on $\scripta$.
\end{itemize}
The importance of the second condition lies in the fact that it
ensures that real classical quantities will be associated with
symmetric operators, which in turn have spectrum on the real line and
real expectation values. If this would not be the case, the
interpretation of the resulting quantum theory would be completely
obscure.
 
The second step consists in choosing a
$*$-representation of the algebra $\scripta$, thus enabling one to
compute expectation values and hence make physical
predictions.

The purpose of this section is to look at the
first of these two steps in the context of LQG. It has been
extensively studied there and the choice of the set of classical
observables as well as the corresponding $*$-algebra which is made can
be regarded as the very essence of LQG.
In this section, we
will briefly review these developments to make the paper self
contained as well as fix the notation. 

As a first step recall that the canonical pair in LQG is a SU(2)
connection one-form $A$ and a frame field $E_I$ with a nontrivial
density weight. Both of these take values on a spacial slice $\Sigma$
of the four-manifold $M$.  Being a one-form, $A$ can be integrated
naturally (that is, without recurse to background structure) along
curves $e$ in $\Sigma$, to form \textit{holonomies}
\begin{equation*}
h_e[A]=\porder\exp\left[i\int_e A_a ds^a \right].
\end{equation*}

It turns out to be convenient to consider functions of $A$ which are 
slightly more general.
\begin{defi}
    A \textit{graph} in $\Sigma$ is a collection of analytic, oriented curves 
    in $\Sigma$ which intersect each other at most in their 
    endpoints.

    A functional $c$ depending on connections $A$ on $\Sigma$ just in
    terms of their holonomies along the edges of a graph, i.e. 
    \begin{equation*}
      c[A]\equiv c(h_{e_1}[A],h_{e_2}[A],\ldots,h_{e_n}[A] ), \qquad
      e_1,e_2,\ldots, e_n\quad\text{ edges of some }\gamma, 
    \end{equation*}
    where $c(g_1,\ldots, e_n)$ viewed as a function on $\sutwo^n$ is
    \textit{continuous}, will be called \textit{cylindrical}. 
\end{defi}
Let us denote the set of all cylindrical functions by $\cyl$. It can
be turned into a *-algebra by defining addition, multiplication, and involution by, respectively, pointwise addition, pointwise multiplication, and pointwise complex conjugation. Note that analyticity of the edges 
insures that cyl is closed under multiplication. 

It turns out that the algebra of cylindrical functions can be equipped with 
a norm (derived from the sup-norm for functions on SU(2)$^n$) such that 
its closure $\overline{\cyl}$ with respect to that norm is a
commutative C$^*$-algebra. We will not spell out the details of this
construction but refer the reader to the beautiful presentations
\cite{Ashtekar:1994mh,Ashtekar:1994wa}. We note furthermore that by
changing the word ``continuous'' 
in the definition of $\cyl$ to ``$n$ times differentiable'', we can define
subsets $\cyl^n$ of $\cyl$ (and hence of $\overline{\cyl}$) and, most 
importantly for us, 
\begin{equation*}
  \cyl^\infty\doteq \bigcap_n\cyl^n, 
\end{equation*}
the space of smooth cylindrical functions. 

The density weight of $E$ on the other hand is such that, using an
additional real (co-)vector field $f^i$, it can be naturally integrated
over oriented surfaces $S$ to form a quantity
\begin{equation*}
E_{S,f}= \int_S E^a_i f^i \epsilon_{abc}\,dx^b\,dx^c
\end{equation*}
analogous to the electric flux through $S$. Let us denote by 
$\E$ the linear space spanned by these variables.

One of the defining choices 
of LQG is to base the quantization precisely on the elements of 
$\overline{\cyl}$ and the fluxes $\E$
as classical observables.
From the Poisson brackets of $A$ and
$E$ one can compute the Poisson brackets for the $c$,
$E_{S,f}$. This step involves taking certain limits since -- a
  priory -- cylindrical functions and the $E_{S,f}$ are too singular for
  the Poisson bracket between them to be well defined. For a
  satisfying resolution of this problem (with the result stated
  in the following) see \cite{Ashtekar:1998ak}.
Call $\gamma$ adapted to $S$ if all $p$ in $S\cap\gamma$ are vertices 
of $\gamma$. 
Let $c$ be a function cylindrical on $\gamma$ and $S$ some analytical 
surface. Without restriction of generality we assume that $\gamma$ is 
adapted to $S$. (There is always a $\gamma'$ that contains
  $\gamma$ such that $\gamma'$ is 
adapted to $S$. A $c$ cylindrical on $\gamma$ is clearly also
cylindrical on $\gamma'$. For details see \cite{Ashtekar:1996eg}.) Then   
\begin{equation*}
  \left\{E_{S,f},c\right\} = \frac{\kappa}{2}\sum_{p\in S\cap\gamma}
  \sum_{e_{p}} w(e_{p})f_{i}X_{e_{p}}^{i}[c]
\end{equation*}
where the second sum is over the edges of $\gamma$ adjacent to $p$, 
\begin{equation*}
    w(e_{p})=\begin{cases}1&\text{ if $e$ lies above $S$}\\
0&\text{ if $e$ is tangential to $S$}\\ 
-1&\text{ if $e$ lies below $S$}\end{cases},
\end{equation*}
and $X^{I}_{e_{p}}$ is the $I$th left-invariant (right-invariant) 
vector field on SU(2) acting on the argument of $c$ corresponding to 
the holonomy $h_{e_{p}}$ if $e_{p}$ is pointing away from (towards) $S$.
$\kappa$ is the coupling constant of gravity. 

Surprisingly, the Poisson brackets of the $E_{S,f}$ among themselves
do not vanish as one would expect for the momentum observables. This
poses two questions: Can one nevertheless give some well defined
``Poisson bracket goes to commutator''-prescription to associate algebra
elements to classical observables? And: Can one understand where this
non-commutativity of the momentum observables come from? As shown in
\cite{Ashtekar:1998ak}, both questions can be answered affirmatively.
We do not want to repeat the discussion of \cite{Ashtekar:1998ak} here
but just give its
result, condensed in a definition of the algebra $\scripta$ on which
the quantum theory will be based, as well as the association of
classical observables to algebra elements. Let 
\begin{equation*}
    X_{S,f}[c]:=\frac{il_P^2}{2}\sum_{p\in S\cap\gamma} \sum_{e_{p}}
    w(e_{p})f_{i}X_{e_{p}}^{i}[c],
\end{equation*}
where we have used the notation introduced above. 
\begin{defi}
  Let $\scripta$ be the quotient of the \emph{free} algebra generated by
  the cylindrical functions
  $\overline{\cyl}$ and the fluxes $E_{S,f}$ (here just treated as symbols indexing algebra elements), and the following relations: 
\begin{itemize}
\item the relations encoding 
the linear structures on $\cyl$ and $\E$, 
\item the relations encoding the product between
cylindrical functions,  
\item the relations
 \begin{equation}
\label{commut}
    \comm{E_{S,f}}{c}=X_{S,f}[c].
  \end{equation}
whenever $c\in\cyl^1$.
\end{itemize}  
  On
  $\scripta$ a $*$-operation is given by the usual complex conjugation
  on $\cyl$ and the trivial $(E_{S,f})^*:=E_{S,f}$.
\end{defi}
A definition of $\scripta$ which states all the relations in $\scripta$ explicitly can be found in \cite{Lewandowski:2005jk}. 

The association of the classical functionals $\cyl, \E$ 
  with elements of $\scripta$ is then given by
\begin{align*}
  c &\mapsto c,\\
  E_{S,f} &\mapsto E_{S,f},\\
  \left\{E_{S,f},E_{S',f'}\right\} &\mapsto i\hbar^{-1}\comm{E_{S,f}}{E_{S',f'}},
\end{align*}
and higher order Poisson brackets of elements of $\E$ are mapped to
the higher order commutators of the corresponding elements of $\scripta$.
 
Note that since $\scripta$ is generated by the elements of 
$\cyl$ and $\E$, a representation $\pi$ of $\scripta$ is completely 
determined once the representors $\rep{\cyl}$ and $\rep{\E}$ are known.  
\section{Remarks on the representation theory of $\scripta$}
In the present section we will make some simple observations on the 
structure of representations of $\scripta$. Before we proceed to the
details, let us give a brief outline of what we are going to do.

We assume that a representation of $\scripta$ is given. As a first
step, we appeal to the  powerful machinery available
for representation of C$^*$-algebras, to decompose
the representation space into subspaces on which $\cyl$ acts cyclic. 
Then we look at the action of the representors of the $E_{S,f}$ with
respect to this decomposition. Since the respective operators will be
unbounded, we have to make an assumption that gets possible domain
problems out of the way: We assume that the $\rep{E_{S,f}}$
share a certain dense set in their domains. Then we make the central 
observation that under this assumption, the action of the 
$\rep{E_{S,f}}$ is rather
simple: Roughly speaking it is the sum of a derivative defined by
$X_{S,f}$ and a multiplication operator.
As a consequence, we can show that each representation of $\scripta$
is uniquely determined by a set of measures and functions on the space 
of (generalized) connections fulfilling certain compatibility
conditions.   
Since despite our assumption the considerations might appear
exceedingly general, we will finish by giving a useful corollary of
our results in a rather simple case. 

Before we start our analysis of the representations of $\scripta$ we
want to recall some basic facts about the representation theory of
$\cyl$. As was realized in \cite{Ashtekar:1991kc,Ashtekar:1994mh}, many powerful
results are at hand because $\cyl$ is an unital Abelian
C$^*$-algebra.
Firstly we recall that, due to a theorem of Gel'fand (see for example
\cite{Bratteli:1979tw}), 
since $\cyl$ is Abelian, it is isomorphic, via some 
isomorphism $\iota$, to the algebra of continuous
functions on the \textit{spectrum} $\overline{\mathcal{A}}$, a compact 
Hausdorff space, of $\cyl$. From this and the Riesz-Markov Theorem (see 
for example \cite{Reed})
it follows that every positive linear functional
on $\cyl$ is given by a positive Baire measure on $\abar$.
The converse trivially holds true: Every positive Baire measure on
$\abar$ gives a positive functional on $\cyl$.

Now let $(\pi,\hilb)$ be a \textit{cyclic} representation of $\cyl$.  
Since the representation is cyclic, it 
defines a positive linear functional $\omega$ on $\cyl$ and certainly 
is unitarily equivalent to the GNS representation coming from $\omega$. 
Moreover as concluded above $\omega$ must be given by a positive Baire 
measure on $\abar$. Vice versa, every cyclic representation of $\cyl$ is given
by a positive Baire measure on $\abar$. 
Thus, we conclude that the cyclic representations of $\cyl$ are all of 
the form 
\begin{equation}
\label{eq7}
\hilb_\nu=L^2(\overline{\mathcal{A}}, d\mu_\nu),\qquad \pi_\nu(c)=\iota(c),
\end{equation} 
where $\mu_\nu$ is some positive Baire measure on $\abar$. 
Note that we did not have to assume continuity of the
cyclic representation. Rather, continuity follows automatically from
cyclicity.

It is important for the rest of this work, that because of their
structure \eqref{eq7}, for cyclic representations 
the $\rep{c}$ play a double role: On the one hand they are
operators on the representation space, on the other hand they are
$L^2$ functions. Let us note the following Lemma which will be
useful, 
later on:
\begin{lemm}
Let $(\pi,\hilb)$ be some cyclic representation of $\cyl$. 
Then $\rep{\cyl^{\infty}}$ is dense in $L^2(\overline{\mathcal{A}},d\mu)$. 
\end{lemm}
\begin{proof}[Sketch of the proof]
Since we do not want to introduce the projective limit machinery that
is used to define the closure $\cyl$ of the set of cylindrical
functions, we will only sketch the proof. The details can however be
easily fixed using the methods of
\cite{Ashtekar:1994mh,Ashtekar:1994wa}.

The idea for the proof is that functions in $\cyl^\infty$ can
essentially be viewed as subset of the continuous functions on a
compact space.    
They are separating points and the constant functions are
among them, so the Stone-Weierstrass Theorem (see for example \cite{Stone}) 
applies, showing that they are dense in $\cyl$ (wrt. its C$^*$ norm). 
Now cyclicity of the representation just means that $\rep{\cyl}$ is
dense in $\hilb$, whence $\rep{\cyl^{\infty}}$ is dense in $\hilb$ as
well. 
\end{proof}
Let now a representation $(\pi,\hilb)$ of $\scripta$ 
be given. It is well known that every representation of a C$^*$-algebra is a
direct sum of cyclic representations (see for example \cite{Conway}). 
Applying this to the representation $\pi\rvert_{\cyl}$ of $\cyl$ yields 
\begin{equation*}
\hilb \isom \bigoplus_\nu \hilb_\nu, \quad \quad \pi \isom \oplus_\nu \pi_\nu. 
\end{equation*}
where the $(\pi_{\nu},\hilb_{\nu})$ are cyclic and therefore 
\begin{equation*}
    \hilb_\nu\isom L^2(\overline{\mathcal{A}}, d\mu_\nu),\qquad 
    \pi_\nu(c)\isom \iota(c).
\end{equation*}   
To simplify notation in what follows, we will take all isometries as 
identities. Furthermore, we denote by $I_{\nu}$ the canonical 
inclusion
\begin{equation*}
    I_{\nu}: \hilb_{\nu}\hookrightarrow\hilb 
\end{equation*}
and by $P_{\nu}$ the canonical projection followed by the inverse of 
$I_{\nu}$
\begin{equation*}
    P_{\nu}:\hilb\rightarrow \hilb_{\nu}.
\end{equation*}
Now we have to analyze the action of the operators $\rep{E_{S,f}}$ on
$\hilb$. This gets complicated by the fact that they represent vector
fields and will therefore be unbounded operators. To get these
complications out of the way, we will make an assumption on $\pi$. To this 
end, let us define the following subspace of $\hilb$:
\begin{equation*}
    \Hilb:=\Span\left[ \bigcup_{\nu}I_{\nu}\left(\cyl^\infty\right)\right].
\end{equation*}
Note that $\Hilb$ is dense in $\hilb$ because $\cyl^\infty$ is dense in 
$\hilb_{\nu}$. 
With this definition at hand, we can state our 
assumption:  
\begin{assu}
  \label{ass2} 
  The representation $\pi$ should be such that
  $\Hilb \subset \dom\left(\rep{E_{S,f}}\right)$ for all surfaces
  $S$ and co-vector fields $f$ on $S$.
\end{assu}
We note that this assumption \textit{does not} automatically 
follow from the perhaps more natural one that
$\{\rep{c}\,|\,c\in\cyl\}$ should be contained in the domains of the
$E_{S,f}$. Under this assumption, the action of the $\rep{E_{S,f}}$ can be 
computed rather explicitly: Let $c$ be a cylindrical function. Then 
\begin{align*}
    \rep{E_{S,f}}\left[I_{\nu}(c)\right] 
    &=\rep{E_{S,f}}\rep{c}\left[I_{\nu}(\one)\right]\\
    &=\comm{\rep{E_{S,f}}}{\rep{c}}\left[I_{\nu}(\one)\right]
    +\rep{c}\rep{E_{S,f}}\left[I_{\nu}(\one)\right]\\
    &=I_{\nu}\left(X_{S,f}[c]\right) + 
    \sum_{\iota}I_{\iota}\left(cF^{\iota\nu}_{S,f}\right)
\end{align*}
where we have made the definition 
\begin{equation*}
    F^{\iota\nu}_{S,f}
    :=P_{\iota}\left(\rep{E_{S,f}}\left[I_{\nu}(\one)\right]\right)
    \in \hilb_{\iota}.
\end{equation*}
Thus the action of the fluxes on $\Hilb$ is 
completely determined by the $F_{S,f}^{\iota\nu}$. (It should be noted that several self-adjoint extensions to $\hilb$ may be possible, a priori.) Let us exhibit some 
further properties of this family:

From 
Assumption \ref{ass2} we get
$c\sum_{\iota}I_{\iota}\left(F^{\iota\nu}_{S,f}\right)\in\hilb$ for
cylindrical functions $c$. Then, since $\cyl$ also contains the constant 
functions, we have that
\begin{equation}
\label{dom}
  \tag{dom}
    \sum_{\iota}I_{\iota}\left(F^{\iota\nu}_{S,f}\right)\in\hilb. 
\end{equation}
Another property that follows is that for a differentiable cylindrical function $c \in \cyl^1$ that 
is null with respect to $\mu_\nu$, also
\begin{equation*}
\label{nul}
\tag{nul}
X_{S,f}[c]=0 \quad \mu_\nu\text{-a.e., } \text{ and }
cF^{\iota\nu}_{S,f}=0 \quad \mu_\iota\text{-a.e.}
\quad \forall \iota\neq\nu
\end{equation*}
More properties come from the fact that the $\rep{E_{S,f}}$ represent 
the $E_{S,f}$. First of all, for co-vector fields $f,f''$ on a 
surface $S$ and $f'$ on $S'$   
\begin{gather}
\label{rep}
\tag{rep}
\begin{split}
    F^{\iota\nu}_{S,f+f''}&=F^{\iota\nu}_{S,f}+F^{\iota\nu}_{S,f''},\\
    F^{\iota\nu}_{S,f}+F^{\iota\nu}_{S',f'}
    &=F^{\iota\nu}_{S\setminus S',f}+F^{\iota\nu}_{S'\setminus S,f'} 
    +F^{\iota\nu}_{S\cap S',f\pm f'},
\end{split}
\end{gather}
where in the second line $S \cap S'$ is given the orientation of $S$
and the sign  
depends on the relative orientation of $S$ and $S'$ on their
intersection.
 
Further relations come from the fact that $\pi$ is a $*$-representation: 
For $c,c'\in\cyl$ let 
\begin{equation*}
  \Delta_{S,f}^{(\iota)}(c,c'):=\scpr{X_{S,f}[c]}{c'}_{\hilb_\iota}-
 \scpr{c}{X_{S,f}[c']}_{\hilb_\iota}
\end{equation*}
denote the divergence of the vector field $X_{S,f}$ with respect to
the measure $\mu_\iota$. We then have
\begin{align*}
    \scpr{\rep{E_{S,f}}I_{\nu}(c)}{I_{\iota}(c')}
    &=\scpr{I_{\nu}(c)}{\rep{E_{S,f}}I_{\iota}(c')}\\
    \Leftrightarrow \delta_{\nu\iota}\Delta^{(\iota)}_{S,f}(c,c')
    &=\scpr{c}{F^{\nu\iota}_{S,f}c'}_{\hilb_{\nu}}
    -\scpr{F^{\iota\nu}_{S,f}c}{c'}_{\hilb_{\iota}}.
\end{align*}
To summarize, we found that 
\begin{equation}
\label{div}
\tag{div}
\begin{split}
    \Delta_{S,f}^{(\iota)}(c,c')
    &=2i\scpr{c}{\im\left(F_{S,f}^{\iota\iota}\right)c'},\\
    F^{\nu\iota}\,d\mu_{\nu}&=\overline{F^{\iota\nu}}\,d\mu_{\iota} 
    \qquad \text{for } \iota\neq \nu.
\end{split}
\end{equation}
Let us summarize our findings 
\begin{prop}
\label{pr1}
Any representation $(\pi,\hilb)$ of $\scripta$, fulfilling our
Assumption \ref{ass2}, determines
\begin{itemize}
  \item A family of positive measures $\{\mu_{\nu}\}$ on $\abar$, 
  \item A family of functions $\{F^{\iota\nu}_{S,f}\}$, where  
$F^{\iota\nu}_{S,f}\in L^2(\abar,d\mu_{\iota})$, 
\end{itemize}  
such that \eqref{dom}, \eqref{nul}, \eqref{rep}, \eqref{div} are fulfilled.  
\end{prop}
It is probably more interesting to note that also the converse holds true:
\begin{prop}
  \label{pr2}
  Let a family of measures $\{\mu_{\nu}\}$ on $\abar$ and a family of 
  functions $\{F^{\iota\nu}_{S,f}\}$, where  
  $F^{\iota\nu}_{S,f}\in L^2(\abar,d\mu_{\nu})$ that fulfill 
  \eqref{dom}, \eqref{nul}, \eqref{rep}, \eqref{div}. 
From this data, one can construct a representation $\pi$ of
$\scripta$ on $\Hilb$ that fulfills the Assumption \ref{ass2}. 
\end{prop}
We note that the proposition only guarantees a representation on the common dense domain $\Hilb$. The flux-opearators are symmetric on this domain, but not necessarily self-adjoint. Whether self-adjoint extensions exist is not a priori clear.
\begin{proof}
The proof is quite obvious: Let $\{\mu_{\nu}\}$,
$\{F^{\iota\nu}_{S,f}\}$ fulfilling \eqref{dom}, \eqref{nul}, \eqref{rep},
\eqref{div} be given. 
The representation space is defined as 
\begin{equation*}
  \hilb \doteq\bigoplus_\nu L^2(\abar, d\mu_\nu)
\end{equation*}
whence
\begin{equation*}
  \rep{c}\oplus_\nu f_\nu \doteq \oplus_\nu c f_\nu, \qquad f_\nu\in
  L^2(\abar, d\mu_\nu). 
\end{equation*}
Now
\begin{equation}
\label{eq5}
  \rep{E_{S,f}}[\oplus_\nu c_\nu]\doteq 
  \oplus_\nu\left(X_{S,f}[c_\nu]+c_\nu\sum_\iota
    F^{\nu\iota}_{S,f}\right),\qquad c_\nu \in \cyl^{\infty} 
\end{equation}
defines operators that are well defined because of \eqref{nul},
do act on on $\Hilb$ because of
\eqref{dom}, hence Assumption \ref{ass2} is fulfilled. 
Moreover they are symmetric because of \eqref{div} and give a
representation of the $E_{S,f}$ since the commutator with the
representors of cylindrical functions is right and
\eqref{rep} holds. 
\end{proof}
The above results may appear exceedingly general. 
Let us therefore reduce consideration to representations in which
$\cyl$ acts cyclic and state the following corollary which is perhaps
closer to applications than the above general results: 
\begin{prop}
  \label{pr3}
  Let a \textit{cyclic} representation $(\pi,\hilb)$ of $\cyl$ be given. 
  Then a necessary and sufficient condition for $\pi$ to be
  extendable to a representation, fulfilling Assumption \ref{ass2}, 
  of the whole $\scripta$ is that the derivations $X_{S,f}$ fulfill condition \eqref{nul}, that they contain, at least, $\cyl^\infty$ in their domain, and that for
  each surface $S$ and co-vector field $f$ on $S$ there exists a
  constant $C_{S,f}$ such that
  \begin{equation}
    \label{eq6}
    \betr{\Delta_{S,f}(c,\one)} \leq C_{S,f}\norm{c}_{\hilb}\qquad\text{ for 
      all }c\in \cyl^{\infty}
  \end{equation}
  where the sesquilinear form $\Delta_{S,f}$ is given by
  \begin{equation*}
    \Delta_{S,f}(c,c')\doteq \scpr{\rep{X_{S,f}[c]}\one}{\rep{c'}\one}_{\hilb}-
    \scpr{\rep{c}\one}{\rep{X_{S,f}[c']}\one}_{\hilb}, \qquad c,c'\in\cyl.
  \end{equation*}
\end{prop}
\begin{proof}
Let us first prove necessity: Let a representation 
$(\tilde{\pi},\hilb)$ of $\scripta$ be given such that
$\tilde{\pi}\rvert_{\cyl}=\pi$. Application of Proposition \ref{pr1}
then yields a measure $\mu$ and a family of functions
$\{F_{S,f}\}$ satisfying \eqref{dom}, \eqref{nul}, \eqref{rep}, \eqref{div}. 
Thus we can finish by noting that \eqref{div}, \eqref{dom} imply
\eqref{eq6}.
  
Sufficiency can be proved by straightforward construction: 
Let a cyclic representation $(\pi,\hilb)$ of $\cyl$, fulfilling the
condition \eqref{eq6}, be given.
Because of cyclicity, $\hilb$ is isomorphic to $L^2(\abar,d\mu)$ for
some positive regular measure $\mu$. Moreover, $\rep{\cyl^\infty}$ is
dense in $\hilb$. Therefore 
the Riesz Representation Theorem (see for example \cite{Reed}) shows that
\eqref{eq6} implies that $\Delta_{S,f}(c,\one)$ is given by an element
$\widetilde{F}_{S,f}$ of $\hilb$,
i.e.\ $\Delta_{S,f}(c,\one)=\langle\rep{c}\one,\widetilde{F}_{S,f}\rangle$.

Now note that because $X_{S,f}$ is a derivation and contains a factor of $i$, 
\begin{align*}
  \Delta_{S,f}(c\overline{c'},\one)
  &=\scpr{\pi(X_{S,f}[c\overline{c'}]\one)}{\one}
  = \scpr{\pi(\overline{c'})\pi(X_{S,f}[c])\one}{\one}
  +\scpr{\pi(c)\pi(X_{S,f}[\overline{c'}])\one}{\one}\\
  &=\scpr{\pi(X_{S,f}[c])\one}{\pi(c')\one}-\scpr{\pi(c)\one}{\pi(X_{S,f}[c'])\one}\\
  &=\Delta_{S,f}(c,c').
\end{align*}
Thus $\Delta_{S,f}(c,c')=\langle\rep{c}\one,\widetilde{F}_{S,f}\rep{c'}\one\rangle$.
Moreover $\re(\widetilde{F}_{S,f})=0$, because
\begin{equation*}
  \Delta_{S,f}(\overline{c},\one)=\scpr{\pi(X_{S,f}[\overline{c}])\one}{\one}
=-\scpr{\one}{\pi(X_{S,f}[c])\one}=-\overline{\Delta_{S,f}(c,\one)}
\end{equation*}
and hence for real $c,c'$
\begin{equation*}
  2\langle\rep{c}\one,\re (\widetilde{F}_{S,f})\rep{c'}\one\rangle=
  \Delta_{S,f}(c,c')+\overline{\Delta_{S,f}(c,c')}=\Delta_{S,f}(cc',\one)
  -\Delta_{S,f}(cc',\one)=0.
\end{equation*}
Set $F_{S,f}\doteq \widetilde{F}_{S,f}/2$. The $F_{S,f}$ fulfill
\eqref{dom}, \eqref{rep} because the $\widetilde{F}_{S,f}$ do. 
Moreover, the $F_{S,f}$ satisfy \eqref{div}. 

Finally, note that for a cylindrical function $c$ that is null
a.e. wrt. $\mu$, we have 
\begin{equation*}
\betr{\Delta_{S,f}(c,c')}=\betr{\scpr{\rep{X_{S,f}[c]}\one}{\rep{c'}\one}}
\leq C_{S,f}\norm{c\overline{c'}}=0
\end{equation*}
for all $c'\in \overline{\cyl}$, whence $X_{S,f}[c]=0$ a.e. wrt. $\mu$
as well. So $\mu$ fulfills \eqref{nul}.

Altogether, we have found data $\mu,\{F_{S,f}\}$
fulfilling \eqref{dom}, \eqref{rep}, \eqref{div}, \eqref{nul}. Hence 
Proposition \ref{pr2} gives the desired (extended) representation. 
\end{proof}
\section{Discussion}
Let us start the discussion of the above results by describing the
simplest case, the AL-representation. In that representation, $\cyl$ 
acts cyclic, the corresponding measure on $\abar$ is the 
Ashtekar-Lewandowski measure $\mu_{\text{AL}}$ constructed in 
\cite{Ashtekar:1994mh} and the $F_{S,f}$ are all equal to zero. 

Next, we remark that Assumption \ref{ass2} precludes the possibility that the
generalized divergences $\Delta_{S,f}$ \textit{are} given by functions which
however \textit{are not} $L^2$. This case does not seem unnatural, 
so it might appear too restrictive to exclude it. Note however that
admitting that case as well would mean that not all of $\cyl$ 
(especially not the constant functions) would be contained in the
domains of the $\rep{E_{S,f}}$. On the other hand the
cylindrical functions are the only ones we have direct control on,
and removing some of them would most likely leave us with a set that
is not dense anymore. Thus it would be extremely difficult to work 
with such more general representations.  

Also, we would like to make some remarks concerning Proposition
\ref{pr3}: As we saw, it is simple to derive that result. It turns out 
to be much more difficult to actually come up with an example for a
measure on $\abar$ fulfilling the condition, other then the
AL-measure. All measures constructed so far, with the remarkable
exception of the AL-measure, violate \eqref{eq6}. The interested
reader is referred to \cite{Sahlmann:2002xv} for a closer investigation of this
subject. 
A class of representations that avoids this problem is the one
obtained by using the AL-measure but having the $F_{S,f}$ real and not 
equal to zero. In such representations, the 
$\rep{E_{S,f}}$ have non vanishing expectation values. 
 
As the last remark showed, this paper merely provides a starting point
for the analysis of the representations of $\scripta$, and much more
difficult and interesting problems remain to be tackled. Some recent results in this direction can be found in \cite{Okolow:2003pk,Sahlmann:2003in,Lewandowski:2005jk,Varadarajan:2007dk,Dziendzikowski:2009rv}.
The present work remains a useful preparation for many tasks related to algebra and representations in LQG. 
\\
\\
\\ 
{\bfseries \Large Acknowledgments}
\\
\\
It is a pleasure to thank Thomas Thiemann for numerous valuable
discussions and suggestions concerning the present work. We are also
grateful to Fotini Markopoulou for many very helpful discussions 
on conceptual issues in quantum gravity. 
Financial support from the Studienstiftung des Deutschen Volkes and
the Max Planck-Institut f\"ur Gravitationsphysik are gratefully
acknowledged. Part of this work was also supported by the Spanish MICINN
project No.\ FIS2008-06078-C03-03.

\end{document}